%% file: main.tex
\documentclass[conference]{IEEEtran}
\IEEEoverridecommandlockouts
\usepackage{amsmath,amssymb,amsfonts}
\usepackage{algorithmic}
\usepackage{graphicx}
\usepackage{textcomp}
\usepackage{xcolor}
\usepackage{subfiles}
\usepackage{graphics}
\usepackage{graphicx}
\usepackage{makecell}
\graphicspath{ {./images/} }
\usepackage{xcolor}
\usepackage{tabularx}

\usepackage[utf8]{inputenc}

\newcommand{\scikit}{\texttt{scikit-learn\/}}
\newcommand{\mlpack}{\texttt{mlpack\/}}
\newcommand{\perf}{\texttt{perf\/}}
\newcommand{\perfmem}{\texttt{perf mem\/}}

\def\BibTeX{{\rm B\kern-.05em{\sc i\kern-.025em b}\kern-.08em
    T\kern-.1667em\lower.7ex\hbox{E}\kern-.125emX}}
\usepackage[maxbibnames=99,
style=alphabetic,
]{biblatex}
\begin{document}
\title{Performance Characterization and Optimizations of Traditional ML Applications\\
}

\author{\IEEEauthorblockN{Harsh Kumar}
\IEEEauthorblockA{\textit{Department of Computer Science \& Automation} \\ 
\textit{Indian Institute of Science, Bangalore}\\
harshkumar@iisc.ac.in}
\and
\IEEEauthorblockN{R. Govindarajan}
\IEEEauthorblockA{\textit{Department of Computer Science \& Automation} \\ 
\textit{Indian Institute of Science, Bangalore}\\
govind@iisc.ac.in}
}


\maketitle

\thispagestyle{plain}
\pagestyle{plain}

\begin{abstract}
Even in the era of Deep Learning based methods, traditional machine learning methods 
with large data sets continue to attract significant attention. 
However, we find an apparent lack of a detailed 
performance characterization of these methods in the context of large training datasets.  

In this work, we study the systems behavior of a number of traditional ML methods as implemented in popular free software libraries/modules
to identify critical performance bottlenecks experienced by these applications.
The performance characterization study reveals several interesting insights on the performance of these applications. 
Then we evaluate the performance benefits of applying some well-known optimizations at the levels of caches and the main memory. More specifically, we test the usefulness of optimizations such as (i) software prefetching to improve cache performance and (ii) data layout and computation reordering optimizations to improve locality in DRAM accesses. 
These optimizations are implemented as modifications to the well-known \scikit\, 
library, 
and hence can be easily leveraged by application programmers. 
We evaluate the impact of the proposed optimizations using a combination of simulation and execution on a real system. 
The software prefetching optimization results in performance benefits varying from 5.2\% - 27.1\% on different ML applications while the data layout and computation reordering approaches yield 6.16\% - 28.0\% performance improvement. 
\end{abstract}

\section{\label{sec:introduction}Introduction}
\subfile{sections/introduction}

\section{\label{sec:methodology}Experiment Methodology}
\subfile{sections/experimental_setup_and_methodology}

\section{\label{sec:perf-char}Performance Characterization}
\subfile{sections/performance_characterization_and_results}

\section{\label{sec:discussion}Analysis of Memory Access Patterns}
\subfile{sections/discussion}

\section{\label{sec:prefetching}Impacts of Prefetching}
\subfile{sections/prefetching}

\section{\label{sec:reordering}Impacts of Data layout and Computation re-ordering}
\subfile{sections/reordering}

\section{\label{sec:related}Related Work}
\subfile{sections/related_work}

\section{\label{sec:concl}Conclusions}
\subfile{sections/conclusion}

\printbibliography

\end{document}

%% file: sections/introduction.tex
In recent years, there has been an ever increasing interest in  machine learning and data science research, thanks to the increasing computational capacity and the availability of huge amounts of user data.  Machine learning methods can be widely categorized into two categories: Deep Learning models and the traditional Machine Learning methods. Deep learning models are characterized by huge computational models such as Neural Networks, CNNs, Language Models, etc. and are known to be both computationally demanding and have large memory footprint \cite{DNN_classical}. Traditional Machine Learning models, on the other hand, use regression models, clustering, decision tree or support vector machines to perform the prediction. 
Another important difference 
is that traditional machine learning methods are explainable whereas  deep learning methods are not \cite{molnar}.  

A recent study \cite{Fotis+19}
noted that  Deep Learning methods only account for less than 20\% of data-science 
workload today and that traditional Machine Learning is still the dominant approach~--- either solely or in some hybrid form, where the outputs of the traditional machine learning methods are fed to the Deep learning models for further processing.

As a consequence, it becomes important to revisit these traditional Machine Learning methods and their bottlenecks in context of ever increasing datasets. 
Several research works \cite{svmd1}, \cite{svmd2}, \cite{dbscan1}, \cite{dbscan4}, \cite{rf1}, \cite{rf3} have been carried out to adapt these traditional algorithms for massive datasets. Distributed and parallel implementations such as Weka-Parallel \cite{parallel_weka} and Spark MLLib \cite{mllib}  
have been developed. However, we find an apparent lack of 
a detailed performance characterization of these 
methods, especially in the context of huge datasets. \\

Several studies perform architectural characterization of data mining algorithms 
\cite{Ozisikyilmaz+06}, \cite{Liu_performanceevaluation}, \cite{ghoting},  \cite{mekkat}.  
A few of the popular traditional machine learning algorithms are a part of their benchmark suite. They measure execution times, cache miss ratios, branch performance, parallelization overheads and speedup with increasing number of threads in their studies. However it does not include any micro-architectural
characterization study. 
Mekkat et al. \cite{mekkat} find the workloads to be memory intensive with little to no temporal locality. 
In \cite{Liu_performanceevaluation}, the authors claim that prefetching can help, but 
neither proposes any specific approach  nor reports any potential benefits of prefetching.
The impact of compiler inserted prefetch instructions on a set of data-mining workloads (NU-MineBench \cite{Ozisikyilmaz+06}) is reported in \cite{mekkat}. 
However, the machine learning workloads in their study are limited to very small datasets. 
Studies in \cite{xie} and \cite{z_jia} 
have focused on studying the system behaviour of big-data analytics workloads. They perform a top-down analysis to measure various performance metrics. \cite{xie} focuses on vectorization on the Intel Xeon Phi machine whereas \cite{z_jia} claims the software stack is the problem. 

All these earlier work either focus on very small datasets or they are only interested in the scalability with the increasing number of threads. Also their selection of workloads does not represent the true state of data-science applications in present days. 

In this paper, we are interested in studying the 
micro-architectural characteristics of widely-used traditional machine learning algorithms (in present day data-science) in the context of large datasets.

We focus mainly on single core and low-end servers (with 4 — 8 cores)
in this study for two reasons.  One,  many data science practitioners (beginners, model developers, and users) frequently train and run their models on these  on low-end servers.  Our performance characterization reveals that the bottlenecks and insights are similar across 1 — 8 cores, although their extent may be different. 

Our selection of workloads is quite broad and covers all the modern day use-cases including Classification, Regression, Clustering and Dimensionality-reduction applications. In our setting, the algorithms are free to run on multiple threads, but we focus our study on the bottlenecks encountered in single thread execution. 
Our performance characterization study reveals several interesting insights to improve the performance of these applications. 
Based on these insights, we implement some well-known optimizations, 
such as software prefetching, data and computation reordering 
and 
evaluate their benefits on the traditional machine learning applications. The software prefetching optimization results in performance benefits varying from 5.2\% - 27.1\% on different ML applications while the data layout and computation reordering approaches yield 6.16\% - 28.0\% performance improvement.
Our optimizations are implemented in \scikit\, and thus allowing users to easily leverage them. 
\\
\\
The rest of the paper is organized as follows. In 
Section~\ref{sec:methodology} we 
describe the workload used and  and the experimental methodology. In Section~\ref{sec:perf-char} 
we present the performance characterization results. This will be followed by a discussion on the reasons behind the bottlenecks in Section~\ref{sec:discussion}. Next, we discuss on the impacts of 
hardware and software-based prefetching methods and the impacts of certain data-layout and computation re-ordering algorithms on the overall performance  in Sections~\ref{sec:prefetching} and~\ref{sec:reordering}. A discussion on  related work is presented in Section~\ref{sec:related}. 
Finally, we conclude the paper in Section~\ref{sec:concl}. 

%% file: sections/experimental_setup_and_methodology.tex
As a first step, we 
select a set of 13  traditional Machine learning workloads which are popularly used in present-day 
data science applications, including Lasso \cite{Lasso} and Ridge \cite{Ridge} regressions, Principal Components Analysis (PCA) \cite{PCA}, Latent Dirichlet Allocation (LDA) \cite{LDA}, Support Vector Machines (SVM) (both linear kernel and RBF kernel) \cite{SVM}, KMeans \cite{KMeans}, Gaussian Mixture Models (GMM) \cite{Spall}, K-Nearest Neighbours (KNN) \cite{KNN}, DBSCAN \cite{DBSCAN}, t-SNE \cite{t_SNE}, Decision Trees \cite{Quinlan}, Random Forests \cite{breiman} and Adaboost \cite{Freund+99}. 
We 
categorize the workloads into three sets based on the nature of subroutines they spend most of their time on. These categories are 
Matrix-algebra based methods, Neighbours-based methods and Tree-based methods. The set of workloads in each category is shown in Table \ref{tab:workload_category}. 

To ensure implementation-independent measurements, we perform measurements on two different implementations of these workloads, one in the \scikit\ 
\cite{scikit-learn} v1.0.1 library and another in the \mlpack~\cite{mlpack2018} v3.4.2 library. The \scikit\ library is the most widely used data science library \cite{Fotis+19} whereas the \mlpack\  website~\cite{mlpack2018} claims it to be the fastest library. Also note that the \mlpack\ \cite{mlpack2018} library does not implement SVM with RBF kernel, LDA and t-SNE algorithms. 

\begin{table}[thb]
    \caption{\label{tab:workload_category}Benchmarks Used and their Categories}
    \begin{center}
    \begin{tabular}{|c|c|}
    \hline
    \textbf{Category} & \textbf{Workload} \\
    \hline
    Matrix-based & Lasso, Ridge, PCA, Linear SVM, SVM-RBF, LDA \\
    Neighbour-based & KMeans, GMM, KNN, DBSCAN, t-SNE \\
    Tree-based & Decision Tree, Random Forests, Adaboost \\
    \hline
    \end{tabular}
    \end{center}
\end{table}

We use  Intel VTune profiler \cite{VTune} for identifying the performance bottleneck and the linux \perf\ tool for measuring various performance monitoring counters. The memory profiling is also done using the Linux \perfmem\ tool. Intel's VTune provides high level bottleneck information in terms of percentage of pipeline slots whereas the \perf\ tools gives actual Performance Monitoring Unit (PMU) counter values which are helpful to understand the nature of the bottlenecks.  

\begin{table}[t]
    \caption{\label{tab:processor_details}Machine Architecture}
    \begin{center}
    \begin{tabular}{|c|c|}
      \hline
      \textbf{Parameters}   &  \textbf{Values}\\
      \hline
      Architecture   &  x86\_64, Intel i7-10700 ice-lake, 2.90 GHz\\
      L1D/I Cache & 256KB/256KB \\
      L2 Cache & 2MB \\
      L3 Cache & 16MB \\
      DRAM & 32 GB \\
      FPU & Available \\
      \hline
    \end{tabular}
    \end{center}
\end{table}

As we are only interested in performance bottlenecks, we make use of dummy datasets of size 10 million rows and 20 features. These datasets are generated using the \textit{datasets} module in the \scikit\  library. The generated datasets are then converted in binary formats -- \texttt{npy} format for \scikit\  and \texttt{bin} format for \mlpack. This avoids the overhead incurred due to reading input text files (and subsequent string to float conversion). 

For each workload, we run the measurements for a minimum of eight hours or five training iterations 
whichever is earlier. The workloads are run on a single core (with n\_jobs parameter set to 1) in isolation. No other user process is run in parallel to the workloads during measurements. Average values of three runs is used for comparison and analysis purposes. Finally, the processor and cache configuration of the system used in our evaluation 
is shown in Table \ref{tab:processor_details}. We used the Ubuntu 20.04 operating system and the underlying BLAS, ATLAS and LAPACK libraries were the default distribution from Netlib \cite{Netlib}. We verified the version of these libraries to be v3.9.0.

%% file: sections/performance_characterization_and_results.tex
In this section, we study the performance characterization of traditional ML applications using VTune and perf tool. First, we present in detail our study on single core systems.  In Section \ref{sec:perf-char-multicore}  we present the results for multi-core servers. 
\subsection{Performance Characterization on Single-core System}
\begin{figure}    \centering
    \includegraphics[scale=0.4]{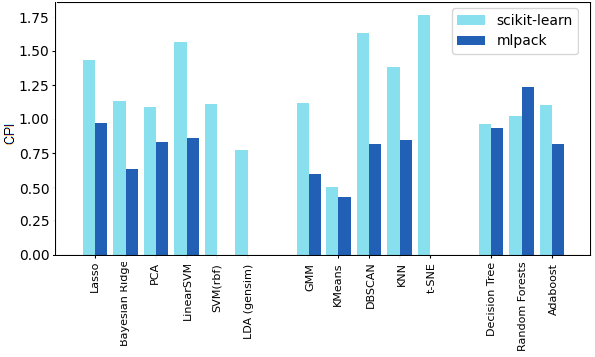}
    \caption{CPI values for all the workloads across both \texttt{scikit-learn} and \texttt{mlpack} implementations.}
    \label{fig:cpi}
\end{figure}

\begin{figure}
    \centering
    \includegraphics[scale=0.4]{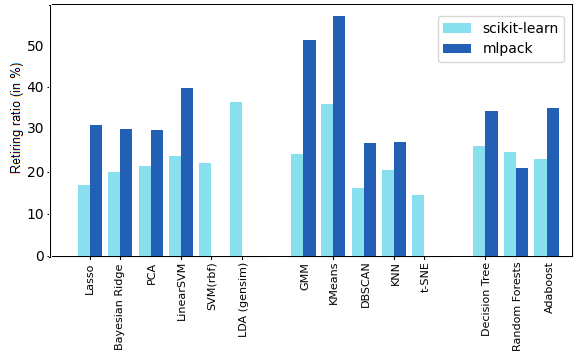}
    \caption{Retiring ratios for all the workloads across both \texttt{scikit-learn} and \texttt{mlpack} implementations.}
    \label{fig:retiring}
\end{figure}

 Figures \ref{fig:cpi} and \ref{fig:retiring} present CPI (Cycles Per Instruction) and retiring ratio (the ratio of instruction slots spent usefully in retiring instructions to the maximum possible instructions that can be retired) expressed as a percentage. 

For all our workloads, the CPI is between 0.4 to 1.75. These CPI values are large given that the processor is an aggressive 5-way 
superscalar architecture. We notice that the 
retiring ratio varies from 15-40\% in all workloads (except GMM and KMeans), indicating potential performance bottlenecks in the applications. Further the performance bottlenecks are higher in \scikit\ implementations than the one using \mlpack.


\begin{figure}
    \centering
    \includegraphics[scale=0.42]{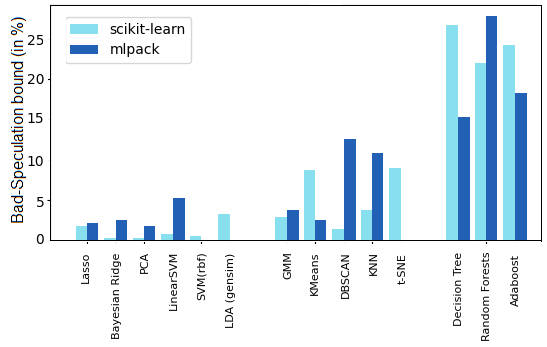}
    \caption{Bad-Speculation bound (percentage) for all the workloads.   }
    \label{fig:bad_speculation_bound}
\end{figure}

\begin{figure}
    \centering
    \includegraphics[scale=0.50]{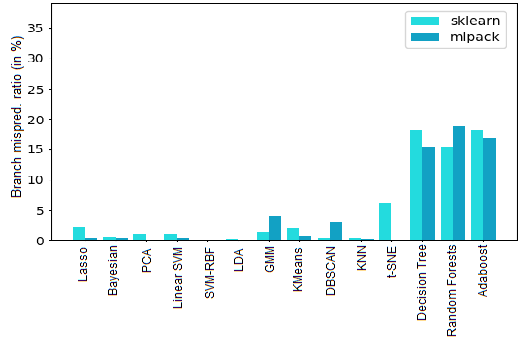}
    \caption{Branch misprediction values for all the workloads across both \texttt{scikit-learn} and \texttt{mlpack} implementations.}
    \label{fig:branch_misprediction}
\end{figure}

\begin{figure}
    \centering
    \includegraphics[scale=0.4]{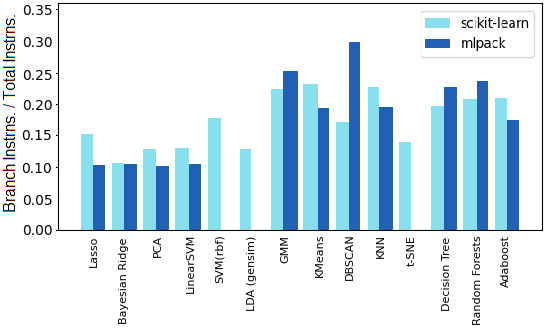}
    \caption{Fraction of branch instructions for all the workloads.}
    \label{fig:fraction_branches}
\end{figure}

\begin{figure}
    \centering
    \includegraphics[scale=0.38]{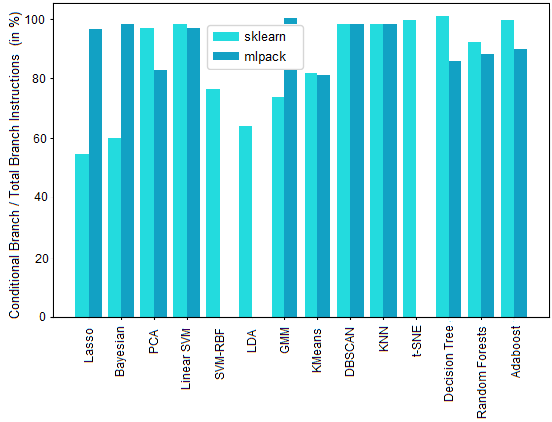}
    \caption{Percentage of conditional branches across all the workloads.}
    \label{fig:fraction_cond_branch_instructions}
\end{figure}

Next to identify the major reasons for the performance bottlenecks, we present the percentage of stalled pipeline slots due to branch mis-speculation in Figure \ref{fig:bad_speculation_bound}. The values are significantly higher in tree-based workloads as seen in Figure \ref{fig:bad_speculation_bound}. Figure \ref{fig:branch_misprediction} shows the corresponding branch misprediction ratios for all the workloads and a positive correlation can be noticed with Figure \ref{fig:bad_speculation_bound}. Further, around 20-25\% of instructions in both neighbour-based and tree-based workloads were found to have branch instructions (see Figure \ref{fig:fraction_branches}). Last, as shown in Figure \ref{fig:fraction_cond_branch_instructions}, around 80-95\% of the branch instruction are conditional branch instructions, implying that the performance of these workloads would be heavily influenced by the quality of the branch-predictor. In neighbour-based workloads, the impact of bad speculation is smaller compared to the tree-based workloads, due to lower misprediction rates (refer to Figure \ref{fig:bad_speculation_bound}).


\begin{figure}
    \centering
    \includegraphics[scale=0.4]{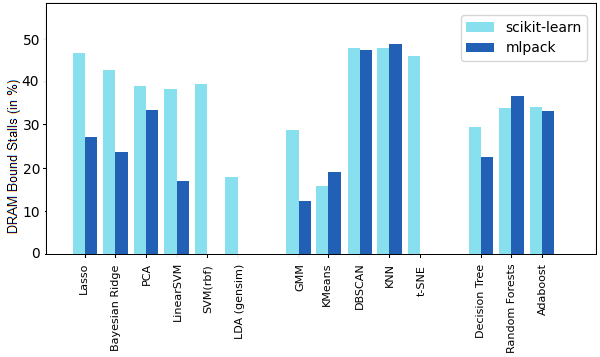}
    \caption{DRAM bound for all the workloads.}
    \label{fig:dram}
\end{figure}

\begin{figure}
    \centering
    \includegraphics[scale=0.4]{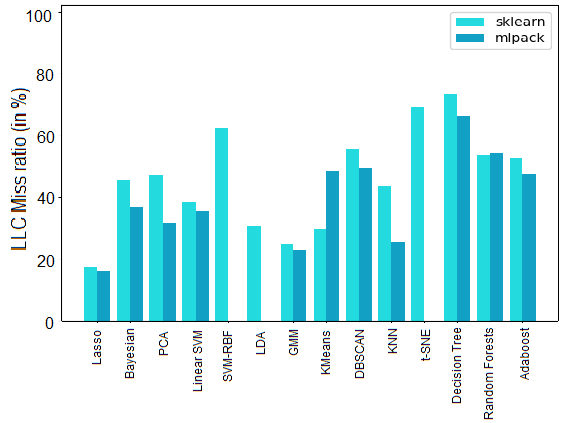}
    \caption{LLC miss ratios for all the workloads.}
    \label{fig:llc}
\end{figure}

Figures \ref{fig:dram} and \ref{fig:llc} respectively show the fraction of cycles stalled due to DRAM (called DRAM bound) and the LLC miss ratio for all the workloads. 
We observe that 31.8\%, 37.4\% and  31.2\% of cycles are stalled 
due to DRAM latency for the matrix-based, neighbour-based and tree based workloads respectively. The LLC miss rates in these workloads also follow a similar trend. These indicate that DRAM latency is one major performance bottleneck in these applications. 


\begin{figure}
    \centering
    \includegraphics[scale=0.42]{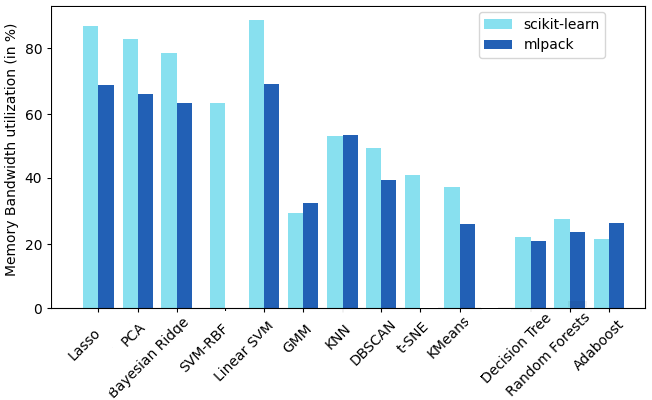}
    \caption{Percentage of memory bandwidth utilization across all the workloads.}
    \label{fig:memory_traffic}
\end{figure}

Figure \ref{fig:memory_traffic} shows the percentage of memory bandwidth utilized by each workload. 
Matrix-based workloads show very high values (around 80\%) of memory bandwidth utilization, whereas other workloads show smaller values (around 40\%). 

\begin{figure}
    \centering
    \includegraphics[scale=0.38]{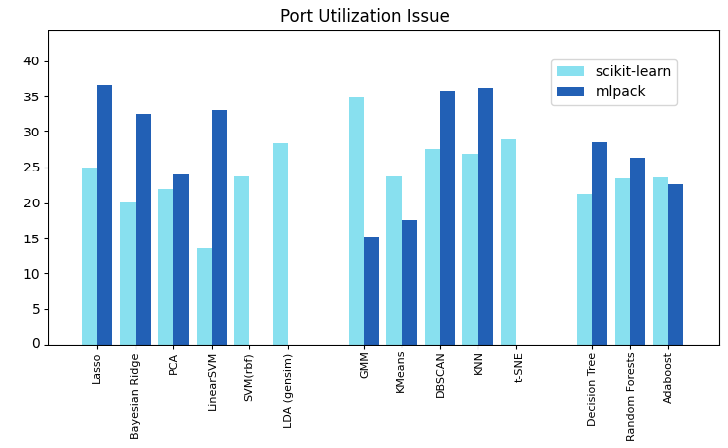}
    \caption{Percentage of execution stall cycles due to core non-divider related issues.}
    \label{fig:port_utilization}
\end{figure}

Last we present port utilization -- the fraction of cycles the CPU execution was stalled on core non-divider related issues -- across all the workloads in Figure \ref{fig:port_utilization}. This again confirms a significant fraction (15\% - 38\%) of the execution cycles are wasted due to functional-unit (other than Divider) related stalls.  

In summary, (i) most workloads have a considerably high CPI (greater than 0.8) and low retiring ratio; (ii) bad-speculation bound caused by high misprediction rate was observed to be a significant performance bottleneck in tree-based workloads (iii) large LLC miss ratios and large fraction of cycles stalled due to DRAM is another major cause of high CPI; and (iv) roughly 25--30\% of the CPU cycles are wasted due to core non-divider related stalls.

\subsection{Multicore Performance Characterization \label{sec:perf-char-multicore}}
Next we present our results for the characterization study on 4 and 8 cores. Due to space constraints we present these results in the form of tables, and limit ourselves only to those benchmarks which have parallel multi-core implementation in the respective library.  For ease of comparison, we have reproduced the 1-core numbers as well in Tables \ref{tab:multicore_sklearn} and \ref{tab:multicore_mlpack}. 

The key observations that we made for single core continue to hold in 4- and 8-core systems.  The CPI’s of the applications continue to be higher (0.7 or higher) and bad speculation and DRAM stalls at 4- and 8-cores are comparable to those at single core.
\begin{table*}
    \caption{Performance numbers for \texttt{scikit-learn} implementation.}
    \centering
    \begin{tabular}{|c|ccc|ccc|ccc|ccc|ccc|}
    \hline
    \textbf{Workloads} & \multicolumn{3}{c|}{\textbf{CPI}} & \multicolumn{3}{c|}{\textbf{Retiring}} & \multicolumn{3}{c|}{\textbf{Bad Spec.}} & \multicolumn{3}{c|}{\textbf{DRAM Bound}} & \multicolumn{3}{c|}{\textbf{Core Bound}}\\
    \hline
    Cores & \textbf{1c} & \textbf{4c} & \textbf{8c} & \textbf{1c} & \textbf{4c} & \textbf{8c} & \textbf{1c} & \textbf{4c} & \textbf{8c} & \textbf{1c} & \textbf{4c} & \textbf{8c} & \textbf{1c} & \textbf{4c} & \textbf{8c}\\
    \hline
        LDA & 0.76& 1.06 & 1.10 & 36.7& 30.3 & 29.8 & 3.2& 4.6 & 3.8 & 19.3 & 31.5 & 34.9 & 28.1 & 29.7 & 27.6 \\
        GMM & 1.08 & 1.14 & 0.85 & 23.5 & 24.2 & 31.0 & 8.9 & 3.6 & 1.1 & 29.5 & 28.7 & 28.6 & 12.4 & 26.0 & 19.8 \\
        KMeans & 0.51 & 1.45 & 0.845 & 37.4 & 24.0 & 35.1 & 8.9 & 6.3 & 9.8 & 15.3 & 16.2 & 18.1 & 19.1 & 16.5 & 21.0 \\
        DBSCAN & 1.61 & 1.64 & 1.32 & 18.1 & 14.6 & 35.3 & 1.9 & 1.4 & 2.5 & 48.5 & 42.7 & 30.3 & 27.4 & 20.6 & 16.7 \\
        kNN & 1.42 & 1.11 & 0.86 & 20.6 & 23.8 & 21.0 & 4.8 & 14.1 & 10.4 & 48.4 & 44.5 & 60.2 & 26.4 & 20.6 & 13.4 \\
        t-SNE & 1.73 & 1.68 & 1.36 & 15.5 & 14.5 & 10.9 & 8.4 & 9.4 & 10.4 & 44.6 & 38.2 & 60.3 & 29.2 & 21.2 & 13.5 \\
        R. Forests & 1.01 & 1.13 & 0.96 & 24.4 & 24.2 & 27.9 & 22.3 & 17.7 & 27.1 & 33.4 & 32.2 & 35.8 & 23.8 & 16.0 & 16.3 \\
        Adaboost & 1.13 & 0.95 & 0.94 & 22.6 & 26.5 & 25.8 & 24.8 & 24.0 & 31.2 & 32.8 & 26.7 & 27.5 & 24.3 & 15.5 & 14.3 \\
        \hline
    \end{tabular}
    \label{tab:multicore_sklearn}
\end{table*}
\begin{table*}
    \caption{Performance numbers for \texttt{mlpack} implementation.}
    \centering
    \begin{tabular}{|c|ccc|ccc|ccc|ccc|ccc|}
    \hline
    \textbf{Workloads} & \multicolumn{3}{c|}{\textbf{CPI}} & \multicolumn{3}{c|}{\textbf{Retiring}} & \multicolumn{3}{c|}{\textbf{Bad Spec.}} & \multicolumn{3}{c|}{\textbf{DRAM Bound}} & \multicolumn{3}{c|}{\textbf{Core Bound}}\\
    \hline
    Cores & \textbf{1c} & \textbf{4c} & \textbf{8c} & \textbf{1c} & \textbf{4c} & \textbf{8c} & \textbf{1c} & \textbf{4c} & \textbf{8c} & \textbf{1c} & \textbf{4c} & \textbf{8c} & \textbf{1c} & \textbf{4c} & \textbf{8c}\\
    \hline
        GMM & 0.62 & 0.56 & 0.68 & 50.8 & 34.5 & 31.6 & 2.9 & 2.7 & 1.1 & 12.4 & 23.8 & 29.1 & 14.9 & 20.4 & 19.8 \\
        KMeans & 0.46 & 0.69 & 0.64 & 56.8 & 41.0 & 39.4 & 2.3 & 1.6 & 1.8 & 19.1 & 16.8 & 20.1 & 17.6 & 14.8 & 17.0 \\
        DBSCAN & 0.78 & 0.98 & 1.32 & 28.1 & 25.8 & 22.9 & 13.5 & 4.5 & 3.6 & 48.4 & 47.5 & 35.2 & 34.3 & 11.5 & 16.4 \\
        kNN & 0.82 & 0.82 & 0.779 & 28.1 & 36.7 & 34.7 & 10.6 & 5.6 & 14.6 & 48.6 & 33.8 & 34.2 & 35.4 & 13.9 & 13.7 \\
        R. Forests & 1.25 & 1.2 & 0.94 & 20.2 & 28.6 & 30.9 & 28.4 & 17.3 & 16.1 & 34.7 & 53.8 & 38.8 & 25.8 & 14.6 & 13.5 \\
        Adaboost & 0.81 & 0.68 & 0.64 & 34.6 & 36.8 & 38.8 & 17.6 & 10.9 & 11.3 & 32.4 & 36.2 & 22.5 & 23.2 & 20.3 & 19.2 \\
        \hline
    \end{tabular}
    \label{tab:multicore_mlpack}
\end{table*}
\\
Next we move on to propose a few standard optimizations to address the key performance bottlenecks observed.  In studying these optimizations and evaluation their impact on performance, we focus on single core system.

%% file: sections/discussion.tex
We saw in Section~\ref{sec:perf-char} that memory accesses is one of the major sources of performance stalls. To understand this better, we did a code-level hotspots analysis using Intel VTune \cite{VTune}. 

For  matrix based workloads, we observed that the memory accesses are regular. We believe the memory access stalls may be due to the inability of the underlying BLAS library to fully reuse the caches. An improvement for these workloads will thus require making changes to the underlying BLAS library. 

\begin{figure}
    \centering
    \includegraphics[scale=0.75]{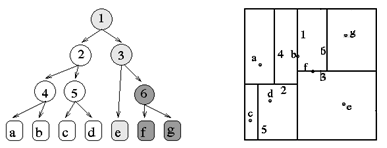}
    \caption{Geometric partition of data samples. In this example, M = 2}
    \label{fig:kdtree}
\end{figure}

For  neighbour-based workloads, however, we noticed the use of irregular memory accesses of the form A[B[i]] (and in some cases, A[B[C[i]]]) to be the main reason behind the bottleneck. It appears that the library implementations use such memory access patterns 
to prune computations. For example, in workloads like KNN and DBSCAN, the neighbourhood information is stored in a tree-based structure (K-D Tree \cite{kdtree} in \texttt{scikit-learn} and Binary Space Tree \cite{binaryspacetree} in \texttt{mlpack}). These tree  structures store the indices of the dataset rows of the samples lying in a certain geometric partition (in the M-dimensional space, where M is the number of features) as shown in Figure \ref{fig:kdtree}. Such an arrangement helps to avoid unnecessary neighbour searches, as the points lying in a specific partition are more likely to be the neighbours. 

For  tree-based workloads, we also noticed the use of irregular memory accesses of the form A[B[i]]. In these workloads, however, the index array B[i] is used to group samples into different nodes of the decision tree. 
We now set out to study the impact of two well-known optimizations, viz., prefetching and memory access ordering on the performance of traditional ML applications.

%% file: sections/prefetching.tex
\newcommand*{\Scale}[2][4]{\scalebox{#1}{$#2$}}
Prefetching reduces  cache misses by fetching the expected data to the caches 
before the demand access happens. The accuracy, coverage and timeliness of the prefetched data is important. Further, care should be taken not to prefetch the data too aggressively or it may lead to cache pollution.

\subsection{\label{sec:pref_benefits}Potential Benefits}
First we measure the potential benefits of prefetching. For this, we measure the improvements in IPC with perfect LLC or perfect L2 cache. We use the Sniper simulator \cite{sniper} for this purpose. The configuration of the simulated system is shown in Table \ref{tab:sniper_config}. 

\begin{table}[tbh]
    \caption{\label{tab:sniper_config}Simulator Configuration}
    \begin{center}
    \begin{tabular}{|c|c|}
      \hline
      \textbf{Parameters}   &  \textbf{Values}\\
      \hline
      Cacheline Size & 64B \\
      L1D/I Cache & 32KB/32KB (8-way) \\
      L2 Cache & 256KB (8-way)\\
      L3 Cache & 8MB (16-way) \\
      Cache replacement & LRU \\
      DRAM & 32 GB, 8 Chips per DIMM, 4 DIMMs \\
      \hline
    \end{tabular}
    \end{center}
\end{table}

\begin{figure}
    \centering
    \includegraphics[scale=0.47]{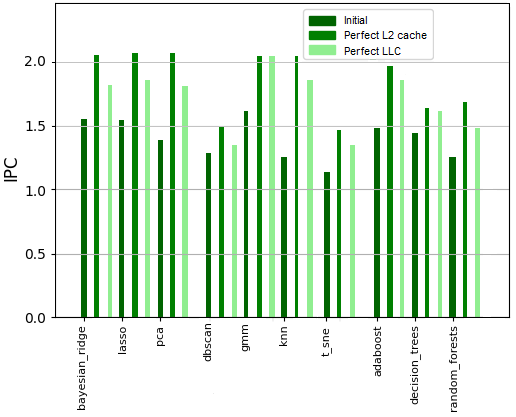}
    \caption{\label{fig:ipc}IPC improvements in case of perfect L2 and LLC for each workload.}
\end{figure}

The results of our simulation study are shown in Figure \ref{fig:ipc}. With perfect LLC, we notice an average improvement of 25.1\%, 35.8\% and 27.1\% for  matrix-based, neighbour-based and tree-based workloads respectively. With perfect L2 cache, we notice an improvement of 40.9\%, 39.22\% and 31.21\% for matrix-based, neighbour-based and tree-based workloads.

\subsection{\label{sec:hwpref}Hardware-based prefetching}
\begin{figure}
    \centering
    \includegraphics[scale=0.38]{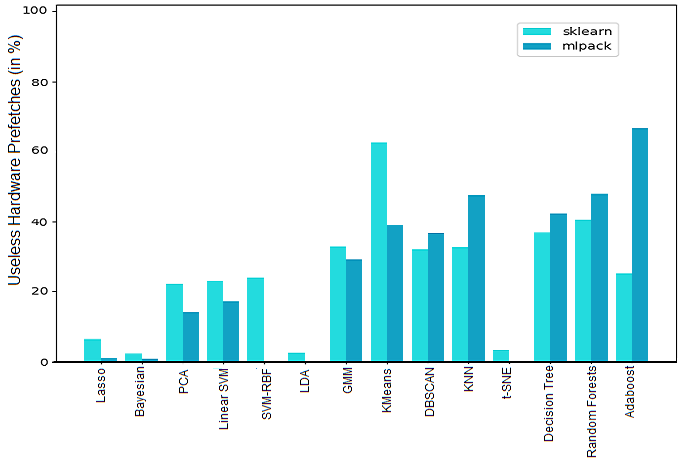}
    \caption{\label{fig:hwpref}Fraction of useless hardware prefetches for each workload.}
\end{figure}

We measure the effectiveness of the default hardware 
in reducing the memory access bottleneck. The fraction of useless hardware prefetches issued is shown in Figure \ref{fig:hwpref}. 

 A major part (nearly 42\%) of the hardware prefetches issued are useless for the neighbour-based and tree-based workloads.  This is due to the irregular nature of the memory access requests.  It may be noted that the performance characterization study reported in Section~\ref{sec:perf-char} did include the benefits of hardware prefetchers, as they are, by default, enabled. 
 
\subsection{\label{sec:swpf}Software-based prefetching}
The memory bandwidth utilization for the Matrix-based workloads already show a very high memory bandwidth utilization. Software-based prefetching in those workloads would only increase the traffic, possibly leading to congestion. We therefore limit our study of software-based prefetching to neighbour-based and tree-based workloads. 

To measure the impacts of the software-based prefetching, we modified the \textit{neighbors} and \textit{tree} modules of the \scikit\, 
library. For this, we used the \texttt{\_mm\_prefetch} intrinsic instruction \cite{intrinsic}. 
These modules are written as Cython \cite{cython} files which first get transformed to a C-language file before finally getting compiled. We manually inserted the intrinsics in the generated intermediate C-language file. In some cases, we also had to unroll a couple of loop iterations to expose opportunities for a timely data prefetch. We targeted the L2 cache as the destination of prefetched data since 
our study in Section \ref{sec:pref_benefits} shows a perfect L2 cache as more promising than the perfect LLC.


\subsection{\label{sec:pref_results}Results}
\begin{figure}
    \centering
    \includegraphics[scale=0.38]{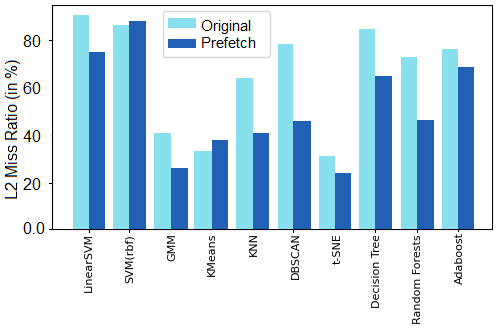}
    \caption{\label{fig:l2_pref}L2 cache miss ratio before and after prefetching}
\end{figure}

\begin{figure}
    \centering
    \includegraphics[scale=0.38]{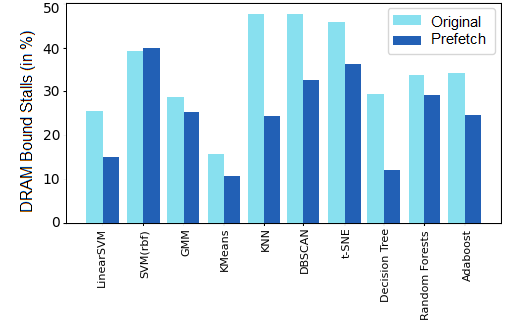}
    \caption{Percentage of DRAM bound stalls before and after prefetching}
    \label{fig:dram_pref}
\end{figure}

We measured the impacts of the software-based prefetching on L2 cache miss rate, DRAM bound, Bad-Speculation bound, port utilization distribution and the resulting performance speedup. Figures \ref{fig:l2_pref} and \ref{fig:dram_pref} show the reduced L2 cache miss rate and DRAM bound. Other than KMeans and SVM, the L2 cache miss rate is reduced by 10\% to 35\%. The DRAM bound stall cycles is reduced by 5\% to 26\% across different workloads. 

\begin{figure}
    \centering
    \includegraphics[scale=0.4]{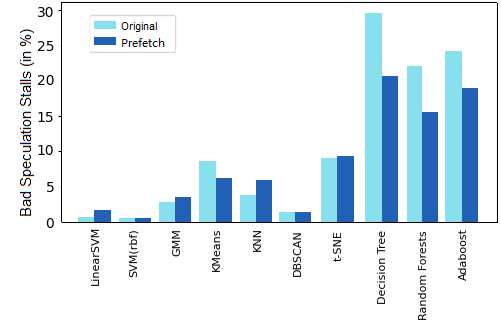}
    \caption{Bad-speculation bound stalls before and after prefetching}
    \label{fig:bad_speculation_pref}
\end{figure}

Interestingly,  bad-speculation bound is also reduced by 8-10\% in tree-based workloads as shown in Figure \ref{fig:bad_speculation_pref}. This could be due to reduced times for branch resolution in cases where the branch result depends on a memory-resident operand. 

\begin{figure}[htb]
    \centering
    \includegraphics[scale=0.25]{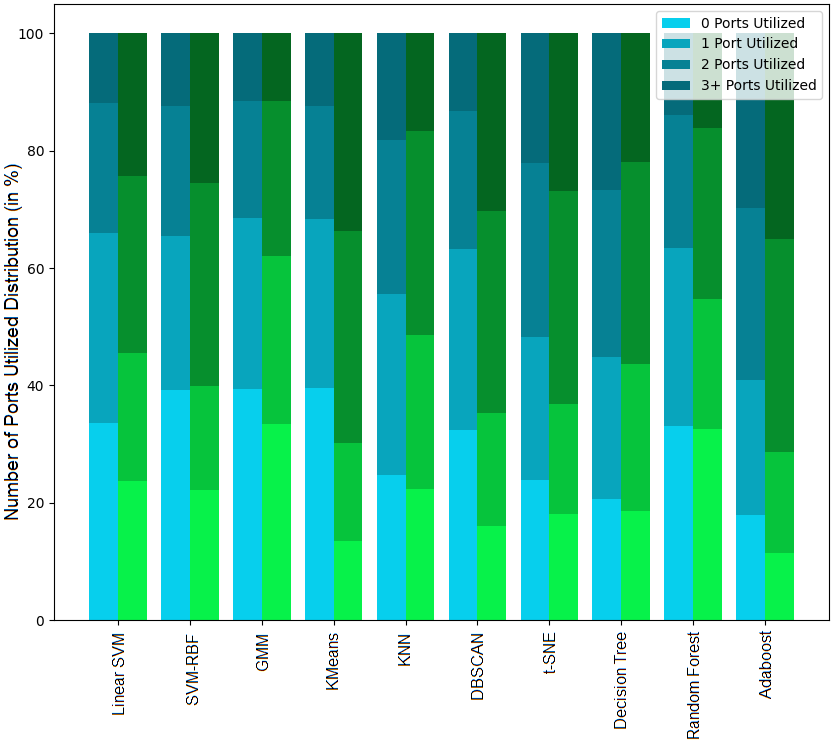}
    \caption{Port utilized distribution before (blue) and after (green) prefetching}
    \label{fig:port_pref}
\end{figure}

\begin{figure}[hbt]
    \centering
    \includegraphics[scale=0.35]{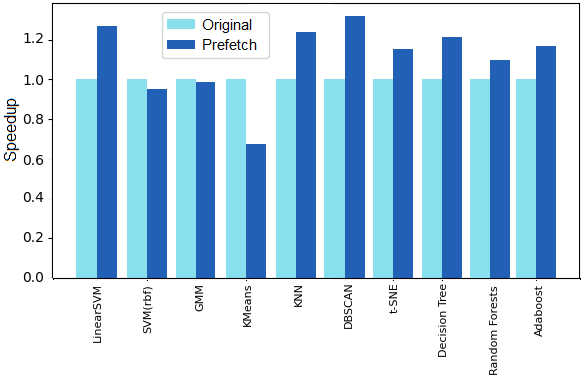}
    \caption{Speedup after prefetching}
    \label{fig:speedup_pref}
\end{figure}

Figure \ref{fig:port_pref} shows an increase of 12.8\% (on average) in the fraction of 2+ uops executed each cycle; this is a further evidence that stalls 
due to dependency on memory-resident operands is reduced by software prefetching. Finally, we show the  speedup achieved by including the software prefetch intrinsics in Figure \ref{fig:speedup_pref}. 
Except in SVM-RBF and KMeans, a speedup of 5\% to 27\% is observed across workloads.  

%% file: sections/reordering.tex
Figures \ref{fig:dram} and \ref{fig:memory_traffic} in Section \ref{sec:perf-char} show a high percentage of DRAM bound stall cycles but with roughly 40\% memory bandwidth utilization for the neighbour-based and tree-based workloads. 
Given the irregular nature of memory accesses in the neighbour and tree based workloads and the large dataset size, 
successive memory accesses are more likely to result in poor spatial locality in both 
cache hierarchy and memory (DRAM). 

\begin{figure}
    \centering
    \includegraphics[scale=0.5]{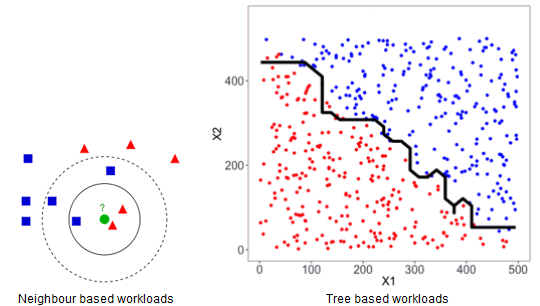}
    \caption{Working of neighbour and tree based workloads}
    \label{fig:geometric_algo}
\end{figure}

As shown in Figure \ref{fig:geometric_algo}, neighbour and tree based methods work on the geometric representation of data samples. Each row in the dataset can be visualized as a point in the M-dimensional space where M is the number of features. 
However, the datasets store the samples as an array of N elements, with the order of samples having no relevance to the proximity in the M-dimensional space. Thus when points which are nearby in the M-dimensional space are accessed in that order, they could be far apart in the dataset rows resulting in poor spatial locality. Further, these accesses are characterized by indirect indexing (e.g., A[B[i]]) and hence cause irregular access pattern.  

Several research works \cite{z_order_computation}, \cite{irregular1}, \cite{irregular2}, \cite{irregular3}, \cite{irregular4}, \cite{irregular5}, \cite{irregular6} have been proposed for improving the performance of irregular workloads. However, most of them deal only with cache optimizations 
and focus on scientific and other applications. To the best of our knowledge we did not find any study in the context of machine learning applications and memory (DRAM) performance.  

\subsection{\label{sec:possible_improvements_reordering}Potential Improvements}
We start with measuring the possible improvements in case of an ideal row-buffer hit ratio. We used the Ramulator \cite{ramulator} DRAM simulator for this purpose. The simulator configuration is shown in Table \ref{tab:ramulator_config}. 
We experimented with two different address mapping schemes: Row-Bank-Rank-Column-Channel and Channel-Rank-Bank-Row-Column, but present results for the former due to space constraints. 
The memory trace was collected using the Linux perf mem tool. 
For our task, we considered only those requests which reach the DRAM (i.e those that are not satisfied by L1/L2/L3 caches). In our study the final memory trace consisted of around 148 million to 2 billion memory accesses. 

\begin{table}
    \caption{\label{tab:ramulator_config}Ramulator Configuration}
    \begin{center}
    \begin{tabular}{|c|c|}
      \hline
      \textbf{Parameters}   &  \textbf{Values}\\
      \hline
      DRAM Standard   &  DDR4\\
      Size & 4GBx8 \\
      Channels & 1 \\
      Ranks & 1 \\
      Banks & 16 per chip \\
      Rows & 32K per bank \\
      Scheduling algorithm & FRFCFS-Cap \cite{frfcfs_cap} \\
      Address scheme & RoBaRaCoCh \\
      \hline
    \end{tabular}
    \end{center}
\end{table}

Columns~2 and~3 in 
Table \ref{tab:dram_improvements} 
report, respectively, the row buffer hit ratio and the average memory access latency experienced by different applications. We observe that 
certain workloads (such as KNN, t-SNE, and DBSCAN) experience a very poor row-buffer hit ratio (less than 0.25), while others have only a moderate 
hit ratio. 
The table also presents (column~4) the
average access latency for an ideal row-buffer hit ratio. The simulation results show a likely improvement of 11.84\% to 25.46\% across the workloads. 
This motivates us to explore different data-layout and computation re-ordering methods that help improve the row-buffer locality.

\begin{table}[hbt]
    \centering
        \caption{\label{tab:dram_improvements}Original and Ideal hit-rate average access latencies for all the workloads.}
    \begin{tabular}{|p{0.22\linewidth}|p{0.12\linewidth}|p{0.12\linewidth}|p{0.12\linewidth}|p{0.17\linewidth}|}
        \hline
        & & & & \\
        \textbf{Benchmark} & \textbf{Original Hit-Ratio} & \textbf{Original Avg. latency} & \textbf{Avg. latency (Ideal Hit-Ratio)} & \textbf{Improvement (\% Memory cycles)} \\
        \hline
        Adaboost & 0.64 & 82.37 & 72.61 & 11.84 \\
        DBSCAN & 0.21 & 91.58 & 69.73 & 23.85 \\
        Decision Tree & 0.36 & 83.06 & 68.14 & 17.96 \\
        GMM & 0.47 & 80.69 & 69.72 & 13.59 \\
        KMeans & 0.48 & 81.58 & 70.71 & 13.33 \\
        KNN & 0.13 & 92.13 & 68.67 & 25.46 \\
        Random Forests & 0.34 & 83.04 & 70.37 & 15.25 \\
        t-SNE & 0.18 & 93.85 & 69.78 & 25.64 \\
        \hline
    \end{tabular}
\end{table}

\subsection{\label{reordering_algorithms}Reordering Algorithms}
In this work, we evaluate five well-known data-layout reordering and computation reordering algorithms, viz., First Touch Reordering \cite{first_touch}, Recursive Co-ordinate Bisection (RCB)\cite{rcb}, reordering based on Space Filling Curves \cite{sfc} 
(Hilbert curve \cite{sfc_book} and Z-order curve \cite{sfc_book}) and Locality based blocking and Z-order based computation reordering \cite{z_order_computation}. 

\begin{table}[hbt]
    \caption{\label{tab:reordering_algo}Reordering algorithms}
    \begin{center}
    \begin{tabular}{|p{0.3\linewidth}|p{0.3\linewidth}|p{0.22\linewidth}|}
      \hline
      \textbf{Category}   &  \textbf{Algorithms} & \textbf{Implementation}\\
      \hline
      First-touch \& RCB data-layout reordering   &  First-touch & Runtime\\ \cline{2-3}
      & RCB & Offline \\
      \hline
      SFC data-layout reordering & Hilbert and Z-order  & Offline \\
      \hline
      Computation reordering & Locality based blocking & Runtime\\ \cline{2-3}
        & Z-order comp. reordering & Runtime\\
      \hline
    \end{tabular}
    \end{center}
\end{table}

\subsection{\label{sec:implementation_reordering}Implementation Details}
The First-touch reordering is based on the inspector-executor model. The inspector inspects the order of data accesses in first iterations. The dataset is then rearranged accordingly. Execution in the subsequent stages proceed in the normal way. 

The Recursive Co-ordinate Bisection, Hilbert curve and Z-order curve reordering algorithms  are based on the geometric representation of the data. Each dataset row is considered as a point in the M dimensional space where M is the number of features. The dataset is thus, a collection of N sample points in the M dimensional space. 
The data set reordering rearranges the elements such that adjacent points in the SFCs occupy adjacent rows if the feature matrix. The reordered dataset is then provided as the input to the workloads. Other the data ordering part,the approach requires no 
change in the implementation libraries is required. 
 
The objective of the Locality based reordering ] is to improve the row-buffer reuse using blocking or tiling. Since the workloads cannot know the virtual address to physical address 
a priori, we limit exploiting row-buffer locality within an OS page. Typically the OS page size is 0.5x or 2x of the row-buffer size. Thus any row-buffer locality exposed within a OS page is likely to be utilized well in the DRAM page (or row-buffer). 
Thus we reorder the memory accesses to ensure that addresses within a virtual page are accessed together.  

The Z-order based computation reordering is implemented using the algorithm described in \cite{z_order_computation}. The First-touch reordering, Locality based blocking and Z-order computation reordering require modifications to the library. We implement these algorithms as modifications to the \texttt{scikit-learn} library. 

\subsection{\label{sec:results_reordering}Results}
\begin{figure}
    \centering
    \includegraphics[scale=0.3]{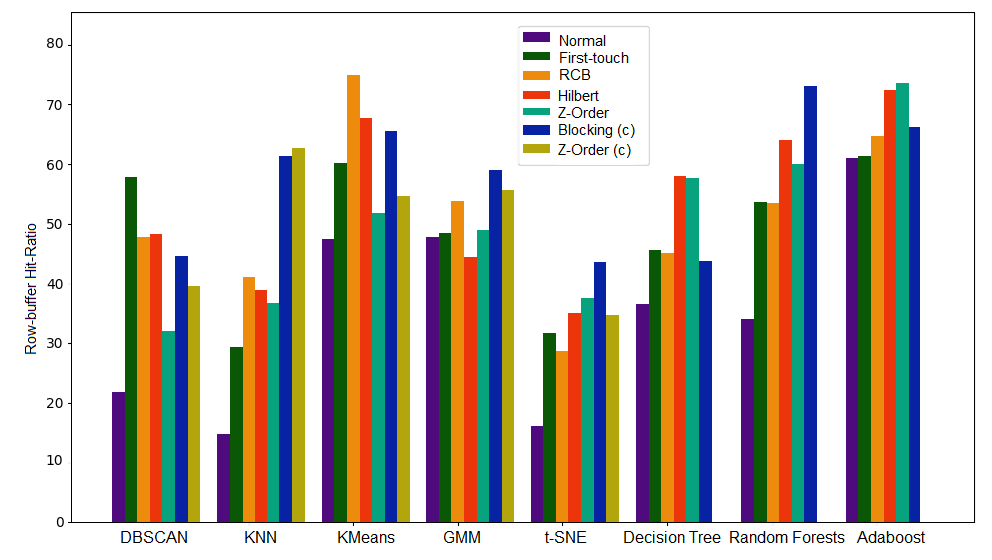}
    \caption{Row-buffer hit ratios for all the workloads. The subscript (c) represents computation reordering.}
    \label{fig:hitrate_1}
\end{figure}

\begin{figure}
    \centering
    \includegraphics[scale=0.4]{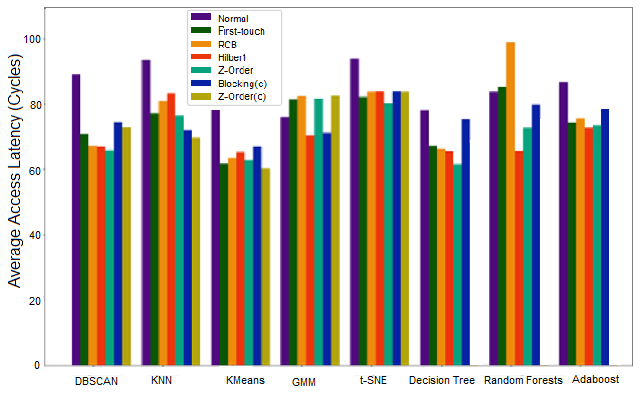}
    \caption{Average access latency for all the workloads. The subscript (c) represents computation reordering.}
    \label{fig:latency_1}
\end{figure}

In this section, we present the impacts of the reordering algorithms. 
We determine, using the Ramulator simulator, the improvements in row-buffer hit ratio and average access latency for our workloads. 
The results are shown in Figures \ref{fig:hitrate_1} and \ref{fig:latency_1}, 
respectively. 

All reordering methods 
improves the hit ratio:  for some applications the improvement is as high as 3x (e.g., First-touch in DBSCAN) or  4X (Z-order(c)  reordering on kNN). Even in applications where the baseline experiences  a good row buffer hit ratio (e.g., Adaboost), reordering approaches give 15\% -- 20\% improvement in hit ratio.
These improvements in row-buffer hit ratio also result in a reduction in the average latency  
by 4.4\%~--~25.14\% across different workloads.  Only in GMM, the average latency increases with 
some of the reordering methods. 


\begin{figure}
    \centering
    \includegraphics[scale=0.35]{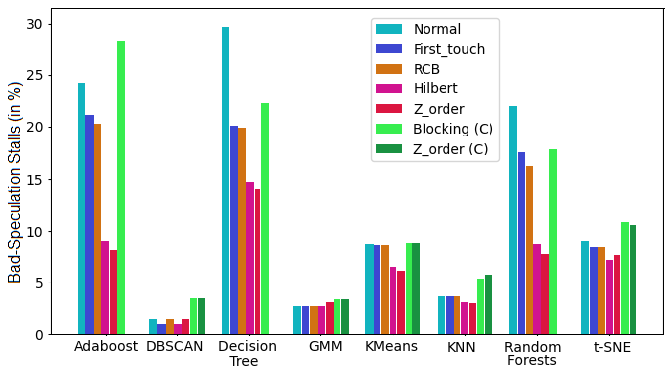}
    \caption{Bad-Speculation bound for all workloads after applying the reordering algorithms.}
    \label{fig:bad_speculation_reordering}
\end{figure}

Significant reductions in bad-speculation stalls for tree-based workloads are observed with maximum reductions for the space filling curves 
(8 to 12\%). As before, we attribute this to early branch resolution. The results are shown in Figure \ref{fig:bad_speculation_reordering}.\\
\\ 

\begin{figure}
    \centering
    \includegraphics[scale=0.35]{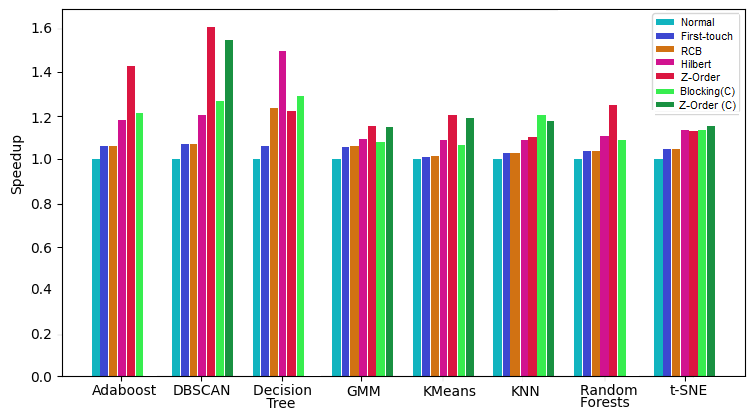}
    \caption{Speedup after applying various reordering algorithms across all
he workloads for 15M rows dataset. No overhead cost considered.}
    \label{fig:speedup_large_no_overhead}
\end{figure}

\begin{figure}[!tbh]
    \centering
    \includegraphics[scale=0.35]{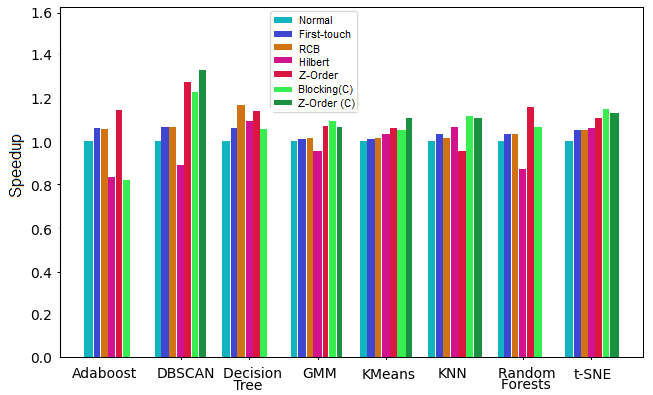}
    \caption{Speedup after applying various reordering algorithms across all
he workloads for 15M rows dataset. Overhead costs of reordering considered.}
    \label{fig:speedup_large_overhead}
\end{figure}

Next, we measured the speedups for each reordering method on actual-runs with 15 million rows dataset.
We noticed a consistent speedup of ranging from 4\% to 60\% across the workloads, when the  reordering overheads are not included (refer to Figure~\ref{fig:speedup_large_no_overhead}).  When the reordering overheads are accounted,  many of the reordering methods have a positive impact on the performance as shown in Figure~\ref{fig:speedup_large_overhead}, with speedup improvement up to 35\%.  Certain reordering methods (such as Hilbert Curve in Adaboost or DBSCAN) incur huge overheads, effectively 
resulting in a slowdown. 


Overall, we observed that the computation reordering algorithms provide relatively better performance improvements for neighbour-based workloads and data-layout reordering algorithms work better for the tree-based workloads, with Random Forest and Adaboost showing best performance improvements after SFC-based data-layout reordering.

\subsection{\label{Comparisons}Comparison of different Reordering algorithms}
In this section, we present a qualitative comparison (refer to Table \ref{tab:comparison}) 
of different reordering methods in term of the overhead incurred and the performance gains achieved for different set of workloads.  

\begin{table}
    \centering
        \caption{\label{tab:comparison}Qualitative comparison of different reordering algorithms}
    \begin{tabular}{|p{0.18\linewidth}|p{0.16\linewidth}|p{0.2\linewidth}|p{0.2\linewidth}|}
        \hline
        \textbf{Category} & \textbf{Algorithm} & \textbf{Neighbour-based workloads} & \textbf{Tree-based workloads} \\
        \hline
        First-touch \& RCB data reordering & First-Touch & Small overheads, medium gains & Small overheads, small gains\\ \cline{2-4}
         & RCB & Small overheads, medium gains & Small overheads, small gains\\
        \hline
        SFC data reordering & Hilbert & Large overheads, high gains & Large overheads, large gains\\ \cline{2-4}
        & Z-Order & Medium overheads, large gains & Same as above\\
        \hline
        Computation reordering & Locality-based Blocking & Medium overheads, medium gains & Medium overheads, small gains \\ \cline{2-4}
        & \makecell[l]{\\Z-Order\\(Index-based)} & Small overheads, medium gains & Not applicable \\
        \hline
    \end{tabular}
\end{table}

Specifically, Hilbert curve and Z-order reordering methods have the largest overheads, 
and advocate the use of these methods only  in cases where the reordered dataset will be used multiple times, for example ensemble based workloads such as Adaboost and Random Forests.
Inspector-executor model based First-Touch reordering has the lowest overheads and can be used in 
other scenarios. 
Further computation re-ordering methods provide the best improvements for the neighbour-based workloads in our experiments. 

As such, we advise using computation reordering algorithms for neighbour-based workloads and data-layout reordering for tree-based workloads.

%% file: sections/related_work.tex

Prior to the modern day data-science, a number of earlier studies have dealt with  data-mining and data-analytics workloads. Both data-science and data-mining fields are related, however, data-mining algorithms mostly deal with finding useful patterns and association rules in a given dataset. A couple of traditional machine learning algorithms studied in our work can also be considered as data-mining workloads. As such, several studies performing data-mining performance characterization such as \cite{Ozisikyilmaz+06}, \cite{Liu_performanceevaluation}, \cite{ghoting}, \cite{bradford} and \cite{mekkat} among others have a few of the traditional machine learning algorithms as a part of their benchmark suite. 

Ozisikyilmaz et al. \cite{Ozisikyilmaz+06}, Liu et al. \cite{Liu_performanceevaluation}, Ghoting et al. \cite{ghoting} and Mekkat et al. \cite{mekkat} performed architectural characterization of data mining applications. Ozisikyilmaz et al. \cite{Ozisikyilmaz+06} also proposed a new data mining benchmark suite called MineBench. As a part of their work, they measured execution times, cache miss ratios, branch performance, parallelization overheads and speedup with increasing number of threads. Mekkat et al. \cite{mekkat} find the workloads to be memory intensive with little to no temporal locality. In \cite{Liu_performanceevaluation}, the authors claim that prefetching can help, but neither propose any specific approach nor report any potential benefits of prefetching. The impact of compiler inserted prefetch instructions on a set of data-mining workloads is reported in \cite{ghoting}. However, 
the compiler-inserted prefetch instructions increase the overall running time in all the machine learning methods in their workload. The dataset size in all these studies is quite small, ranging from 30MB to 65MB. 

Xie et al. \cite{xie} and Jia et al. \cite{z_jia} among others have focused on the system behaviour in context of big-data analytics workloads. They perform a top-down analysis to measure various performance metrics including cache misses, branch mispredictions, Front-end, back-end bound stalls, CPI and speedup. \cite{xie} focus mainly on the extent and impacts of vectorization on Intel Xeon Phi machine. \cite{z_jia} on the other hand find correlations between CPI and other bound stall values. They claim the software stack to be the problem for larger CPI and stall values. We find that KMeans, SVM and PCA appear as a part of their benchmark suite.  

Studies in \cite{awan_memory} and \cite{awan_node} perform top-down analysis using Intel VTune for architectural characterization. \cite{awan_memory} focuses only on measuring  stall values corresponding to different bottlenecks (bounds) whereas \cite{awan_node} measures the impact of the NUMA nodes and next-line hardware prefetchers on execution times. Dimitrov et al. \cite{dimitrov_memory} on the other hand analyze the spatial and temporal reference patterns in physical memory traces at DIMM level. They also study the impacts of caching, prediction and prefetching on these workloads. However, none of our workloads are present in their benchmark suite. 

There have been significant research works also on developing  parallel and distributed frameworks for machine learning workloads. Meng et al. \cite{mllib} propose a distributed ML library based on spark. Ye et al. \cite{vhadoop} proposed a similar library based on Hadoop virtual cluster.  Low et al. \cite{graphlab} propose a framework based on MapReduce to parallelize common patterns in ML algorithms. Finally,  several studies target a reformulation of the algorithms themselves to make them  parallel and distributed. Such studies include \cite{svmd1}, \cite{svmd2}, \cite{svmd3} for SVM, \cite{dbscan1}, \cite{dbscan2}, \cite{dbscan3}, \cite{dbscan4} for DBSCAN and \cite{rf1}, \cite{rf2}, \cite{rf3} for Random Forests among many others.  

In our work, we study a broad set of workloads spanning across  classification, regression, clustering and dimensionality-reduction with much larger datasets, which seems to be missing from the earlier studies. Our set of workloads represents the true state of modern day data-science applications. We expect our study to be useful to other research works focused on developing faster versions of these workloads.

%% file: sections/conclusion.tex
In this work, we carried out  a general performance characterization study on a set of traditional machine learning workloads 
in the context of larger datasets. We found memory access to be a common bottleneck. Tree based workloads also suffer from large bad-speculation stalls. We then evaluated the benefits of two well-known optimizations~--- prefetching and data-layout and computation reordering~--- on the given workloads. We demonstrate significant improvements due these optimizations. Our performance characterization study was done on low-end server systems
with up to 8-cores.  We anticipate the performance characteristics and  bottlenecks 
will continue to hold even in large scale systems, when problem sizes are scaled
proportionately. The   optimizations applied to address the bottlenecks are independent of the number of processor cores.  Hence should work well across multiple cores and high-end servers as well. 